\documentclass[prb,amsmath,twocolumn,amssymb,titlepage]{revtex4-1}

\usepackage{graphicx}
\usepackage{nicefrac}
\usepackage{amsfonts}
\usepackage{amssymb}
\usepackage{amsmath} 
 \usepackage{subfigure}
\usepackage{multirow} 
 \usepackage{tabularx} 
\usepackage{array}
 \usepackage{units}

\usepackage{tensor} 
\usepackage{braket}
\usepackage{mathtools}
\usepackage{bm}
\usepackage{hyperref}

\usepackage{breakurl}
\usepackage{color}
\definecolor{mygray}{gray}{0.6}

\newcommand{\tr}{\operatorname{{\mathrm tr}}} 

\newcommand{\hc}{^{\dagger}}
\newcommand{\ad}{\operatorname{{\mathrm ad}}}
\newcommand{\adn}{\ad_{\hat n}}
\newcommand{\adx}{\ad_{\hat x}}
\newcommand{\adX}[1][]{\ad_{\hat {\mathbf{X}}_{#1}}}
\newcommand{\adp}{\ad_{\hat p}}
\newcommand{\calD}{\mathcal{D}}    
\newcommand{\inbk}[1]{\left[ #1 \right]}
\newcommand{\inbr}[1]{\left\{ #1 \right\}}
\newcommand{\inp}[1]{\left( #1 \right)}

\begin{document}
\title{Entanglement dynamics in a non-Markovian environment: an exactly solvable model}
\author{Justin H.\ \surname{Wilson}}
\affiliation{Joint Quantum Institute and Condensed Matter Theory
  Center, Department of Physics, University of Maryland, College Park,
  Maryland 20742-4111, USA} 
\author{Benjamin M.\ \surname{Fregoso}}
\affiliation{Joint Quantum Institute and Condensed Matter Theory
  Center, Department of Physics, University of Maryland, College Park,
  Maryland 20742-4111, USA} 
\author{Victor M.\ \surname{Galitski}}
\affiliation{Joint Quantum Institute and Condensed Matter Theory
  Center, Department of Physics, University of Maryland, College Park,
  Maryland 20742-4111, USA} 

\begin{abstract}
  We study the non-Markovian effects on the dynamics of entanglement
  in an exactly-solvable model that involves two independent
  oscillators each coupled to its own stochastic noise source. First,
  we develop Lie algebraic and functional integral methods to find an
  exact solution to the single-oscillator problem which includes an
  analytic expression for the density matrix and the complete
  statistics, i.e., the probability distribution functions for
  observables. For long bath time-correlations, we see non-monotonic
  evolution of the uncertainties in observables. Further, we extend
  this exact solution to the two-particle problem and find the
  dynamics of entanglement in a subspace. We find the phenomena of
  `sudden death' and `rebirth' of entanglement. Interestingly, all
  memory effects enter via the functional form of the energy and hence
  the time of death and rebirth is controlled by the amount of noisy
  energy added into each oscillator. If this energy increases above
  (decreases below) a threshold, we obtain sudden death (rebirth) of
  entanglement.
\end{abstract}

\maketitle

\section{Introduction}
\label{sec:introduction}

Noise in quantum systems can lead to abrupt and complete destruction
(sudden death) of entanglement \cite{Yu2004,*Yu2006}. This represents
one of the major obstacles towards building a practical quantum
computer; see for example \cite{DiVincenzo1995}. In particular, when
the bath is Markovian (memoryless), the destruction of entanglement
can be rather swift since the memory of the system's quantum state is
wiped away by its totally uncorrelated interactions with the bath.

Entanglement dynamics including sudden death and birth has been
studied theoretically, e.g., in two-qubit systems in several contexts
\cite{Yu2004,*Yu2006,Yu2007,*Bellomo2007,*Cheng-Li2011,*Diosi2003,*scheel-2003-50,Juan2012,Yonac2006,*Yonac2007}
and in harmonic oscillators
\cite{Paz2008,*Liu2007,*An2009,*An2007,Prauzner-Bechcicki2004}.  The
recent observation of these phenomena in photonic systems
\cite{M.P.Almeida2007} and ensembles of atoms \cite{Laurat2007} has
attracted great interest. In particular, it has been suspected that
bath memory effects could not only provide an avenue to prolong
entanglement but could also lead to its rebirth after it has
experienced sudden death \cite{Bellomo2007}.  However, most noisy
environments are hard to treat analytically by standard techniques
\cite{Breuer2002} and one must use numerics or impose approximations
to obtain a tractable result.
\begin{figure}
\subfigure{\includegraphics[width=0.48\textwidth]{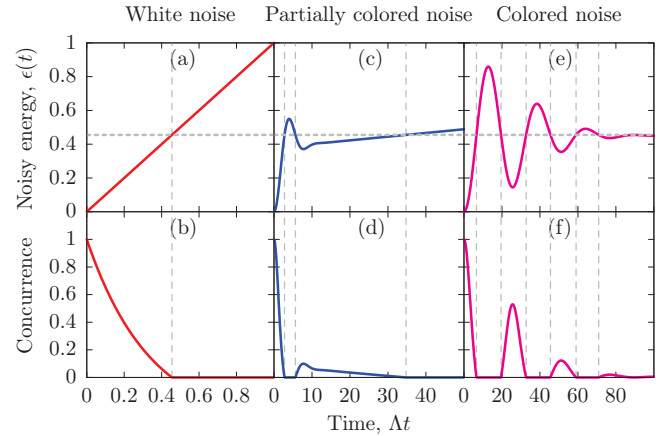}}
\caption{(Color online) The figure shows a comparison between the
  noisy energy of one of the oscillators
  [Eq.~\eqref{eqn:energygaussnoise}] and concurrence (entanglement)
  for noise with different memory. The initial state is
  $(\ket{01}+\ket{10})/\sqrt{2}$.  When the energy exceeds (falls
  below) the threshold $0.455\,\omega$, there is sudden death
  (rebirth) of entanglement. (a,b) use $\Lambda\tau = 0$; (c,d)
  $\omega\tau = 3.5$ and $\omega/\Lambda = 0.875$; (e,f) $\omega \tau
  = 7.5$ and $\omega/\Lambda = 0.25$.}
\label{fig:energy-concurrence-plot-final.eps}
\end{figure}

In this work, we present an exactly solvable model involving two
independent harmonic oscillators each interacting with its own
classical non-Markovian stochastic reservoir.  No back-reaction to the
reservoirs is considered. This system has the property that it can be
solved analytically allowing us to study non-Markovian effects on the
dynamics of entanglement including the prolonging of entanglement and
its rebirth. Particularly, we study the dynamics of entanglement for
the lowest two states of the oscillators which form a qubit-like
system.  Curiously, there is a one-to-one correspondence between the
amount of energy added to each oscillator from the noise source and their
entanglement: As the energy increases (decreases) across a threshold,
we see sudden death (rebirth) of entanglement (see
Fig.~\ref{fig:energy-concurrence-plot-final.eps}). Furthermore, this
initial-state dependent threshold is \emph{independent} of the form of
the noise correlations in time because all memory effects enter via
the energy of a single oscillator which in turn encodes 
the memory effects.

Entanglement between harmonic oscillators can be quantified in several ways
\cite{Prauzner-Bechcicki2004} and can be produced on demand with
trapped ion systems \cite{Turchette1998}. Here we focus on the
lowest two states of each oscillator which form a two qubit-like Hilbert
subspace.  For a two qubit-like system, entanglement is unambiguously
quantified in terms of the \emph{concurrence} $C(\hat{\varrho}_2(t))$,
where $\hat{\varrho}_2$ is the density matrix of two qubit system, we
have
\begin{align}
C(t) = \max\{ 0 , \sqrt{\lambda_1} - \sqrt{\lambda_2} -\sqrt{\lambda_3}-\sqrt{\lambda_4} \},
\label{eq:concurrencedef}
\end{align}
where $\lambda_i$ are the eigenvalues (in decreasing order) of the
matrix $\hat \varrho_2(t) \tilde \varrho_2(t)$ where $\tilde \varrho_2 =
(\sigma_y \otimes \sigma_y)\hat \varrho_2^* (\sigma_y \otimes \sigma_y)$.
Physically, it can be shown \cite{Wootters1998} that states are
maximally entangled if $C(t) =1$ and completely disentangled for
$C(t)=0$. When $C(t)=0$ there exists a realization of $\hat \varrho_2(t)$
such that $ \hat \varrho_2(t) = \sum_k p_k \ket{\psi_k}\bra{\psi_k}$
where every $\ket{\psi_k}$ is separable; i.e., the system is a
classical mixture of separable states. The concurrence can vanish or
appear suddenly at a finite time, counter to what one may naively
expect from the exponential decay of coherences (with characteristic
time $T_2$) which are local quantum phenomena.

In the course of our analysis we first develop the tools to compute
the noise-average density matrix for a single oscillator in the
presence of non-Markovian drive. In addition, we calculate the
probability distribution functions (PDFs) of position, momentum, and
energy observables -- completely characterizing the non-Markovian
statistics of such a system. 

\begin{figure}
\centering
\subfigure{\includegraphics[width=0.48\textwidth]{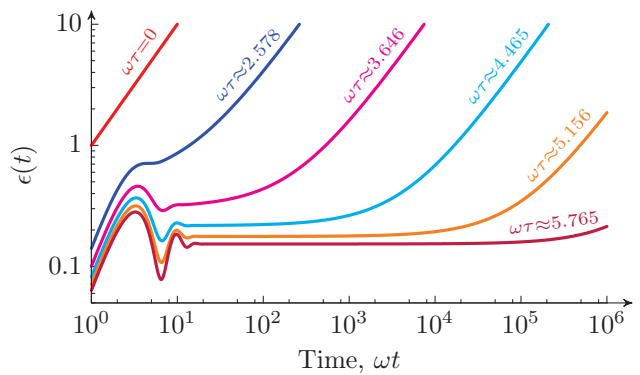}}
\caption{(Color online) The plot is the function
  $\epsilon(t)$ which appears in the energy of the oscillator
  [Eq.~\eqref{eqn:energygaussnoise}], the variances in $\hat x$ and
  $\hat p$ [Eq.~\eqref{eq:variances_x_and_p}], the probability
  distribution functions of position, momentum
  [Eq.~\eqref{eqn:sigma_pdf_new}], and energy
  [Eq.~\eqref{eq:pdfenergy}], and the density matrix
  [Eq.~\eqref{eq:densitymatrix}]. This plot uses $\omega=\Lambda$. }
\label{fig:2}
\end{figure}

In Section~\ref{eq:singleho} we introduce the system and notation, and
we calculate some basic quantities including correlation functions and
energy. In particular, the energy added to the system by the bath
$\epsilon(t)$ (see Fig.~\ref{fig:2}) controls all memory effects that
show up in all later parts of the analysis (including concurrence, as
illustrated in Fig.~\ref{fig:energy-concurrence-plot-final.eps}).  In
Section~\ref{sec:noise-aver-dens} we \emph{analytically} compute the
noise-averaged density matrix (Eq.~\eqref{eq:densitymatrix}) for a
single oscillator in the presence of non-Markovian noise using a
combination of functional integral and Lie algebraic techniques. In
Section~\ref{sec:pdfs}, we calculate the PDFs of position, momentum,
and energy. We find Gaussian PDFs for position and momentum and an
exponential PDF for energy. These PDFs are intimately controlled by
$\epsilon(t)$; they can even contract back towards a delta function
for finite intervals of time before spreading in a diffusive
behavior. In Section~\ref{sec:entanglement} we study the evolution of
concurrence for two oscillators initially maximally entangled (see
Eq.~\eqref{eq:4}) in the subspace of their two lowest states. The
oscillators are independent and subject to independent sources of
non-Markovian noise. We apply the machinery developed in
Section~\ref{eq:singleho} and find an analytical expression for the
effective two-qubit-like density matrix (Eq.~\eqref{eq:8}) used to
calculate the concurrence. We conclude in
Section~\ref{sec:conclusions} with a summary of the main results
derived in this work shown explicitly in Table~\ref{table:summary}.

\section{Single oscillator statistics}
\label{eq:singleho}
In order to study the statistics of a single oscillator, we first
define our system and calculate some basic quantities before moving
onto the bulk of the calculations in Section~\ref{sec:noise-aver-dens}
and \ref{sec:pdfs}. In particular, the energy added to the system by
noise will be important in much of our analysis. The results of this
section are extended to the problem of entanglement of two oscillators
in Section~\ref{sec:entanglement}. 

Our system is characterized by the Hamiltonian of a single driven
harmonic oscillator ($\hbar=1$)
\begin{align}
  \label{eq:3} 
  \hat H = \omega(a^\dagger a + \tfrac12) + \tfrac1{\sqrt2}[\xi (t) a^\dagger + \mathrm{h.c.}],
\end{align}
where $a^{\dagger}(a)$ are the standard creation (annihilation)
operators with $[a ,a^{\dagger}]=1$ and $\xi(t) = \xi_1(t) + i
\xi_2(t)$ defines our external stochastic noise $\xi_{1,2}$ which are
turned on after $t=0$.  The
stochastic forcing terms are completely characterized by their mean
$\langle \xi_i(t) \rangle_{\xi}=0$ and two-time correlation functions
\begin{align}
  \langle \xi_{i}(t) \xi_j(t') \rangle_\xi & = 
  K_{ij}(t,t') =\delta_{ij} k(t-t') \nonumber \\ &
  =\delta_{ij} \tfrac{\Lambda }{\tau
    \sqrt{2\pi}}\,\mathrm{e}^{-(t-t')^2/2\tau^2},
  \label{eq:6}
\end{align}
Our analytical results do not depend on the explicit functional form
of the correlation function $k(t-t')$, but plots and physical
explanations will use the Gaussian time correlations with amplitude
$\Lambda$ and time-correlations $\tau$.  For $\tau=0$, the noise has
no memory and this leads to well known Markovian behavior
\cite{Breuer2002}. We are mostly concerned with the regime where
$\tau\neq 0$.  The average over noise is defined as the functional
integral,
\begin{align}
  \braket{(\cdots)}_{\xi} & = \frac {\int \calD^2 \xi \, (\cdots) \mathrm{e}^{-\frac12 \int_0^t d t' \int_0^t d t'' \xi_i(t')
      K^{-1}_{ij}(t',t'') \xi_j(t'')}}{\int \calD^2 \xi \, \mathrm{e}^{-\frac12 \int_0^t d t' \int_0^t d t'' \xi_i(t')
    K^{-1}_{ij}(t',t'') \xi_j(t'')}},
\end{align}
(summing over repeated indices) where $K^{-1}$ represents the inverse
integral kernel of $K$.

We define the standard occupation number $\hat n=a\hc a$, position  $\hat
x =(a + a\hc)/\sqrt{2}$ and momentum $\hat p =(a
- a\hc)/(\sqrt{2} i)$ operators. The  matrix $R$ is a $2\times 2$
rotation matrix
\begin{align}
  R(t) =
  \begin{pmatrix}
    \cos \omega t & \sin \omega t \\ - \sin \omega t & \cos \omega t
  \end{pmatrix}.
\end{align}
We first study non-Markovian effects in the correlation functions of
position and momentum. The equation of motion for the position and
momentum operators in the Heisenberg picture are $\partial_t \hat
x(t)= \omega \hat p (t) + \xi_2(t)$ and $\partial_t \hat p(t) =
-\omega \hat x(t) - \xi_1(t)$.  Define $\hat{\mathbf{V}}(t) = (-\hat
p(t), \hat x(t))^T$ as a two component vector then solutions can be
written as
\begin{align}
\hat {\mathbf{V}}(t) = R(t) \hat {\mathbf{V}}(0) +
\int_{\mathrlap{0}}^{\mathrlap{t}} ds \,R(t-s)  \bm \xi (s)
\label{eq:classical_eom}
\end{align}
where $\bm \xi(t) = (\xi_1(t),\xi_2(t))^T$ is the external drive.
With these definitions and assuming that the oscillator is initially in a 
number state $\ket n$ the noise-averaged correlation functions are
\begin{multline}
  \langle \langle \hat {\mathbf{V}}(t) \hat {\mathbf{V}}^T(t') \rangle
  \rangle_\xi =
  R(t) \langle \hat {\mathbf{V}}(0)\hat {\mathbf{V}}^T(0) \rangle R(-t')\\
  +\int_{\mathrlap{0}}^{\mathrlap{t}} ds
  \int_{\mathrlap{0}}^{\mathrlap{t'}} ds'\, R(t-s) K(s,s')
  R(s'-t').
\label{eqn:gen_correlationn}
\end{multline}
where $\left\langle \cdots \right\rangle$ is the quantum mechanical
expectation value and $\langle \cdots \rangle_\xi$ is the average over
noise. In particular, from Eq.~\eqref{eqn:gen_correlationn} 
the average of the energy is $\langle \langle \hat{E}(t)
\rangle\rangle_\xi =\omega\langle\langle \hat x^2(t) + \hat
p^2(t)\rangle\rangle_\xi/2= \omega\langle \langle
\tr[\hat{\mathbf{V}}(t)\hat{\mathbf{V}}^T(t)]\rangle\rangle_\xi/2$. Defining 
the energy added to the system due to noise as  
$\omega \epsilon(t)=\langle \langle \hat{E}(t)\rangle\rangle_\xi -\langle \langle \hat{E}(0)\rangle\rangle_\xi=\langle \langle \hat{E}(t)\rangle\rangle_\xi-\omega (n+1/2)$ we find
\begin{align} 
  \epsilon(t) &=\frac{1}{2} \int_{\mathrlap{0}}^{\mathrlap{t}} ds 
  \int_{\mathrlap{0}}^{\mathrlap{t}} ds'\, \tr \{ R(t-s) K(s,s')R(s'-t)\} \\
  &= \int_{\mathrlap{0}}^{\mathrlap{t}}ds
  \int_{\mathrlap{0}}^{\mathrlap{t}} ds'\, \cos\omega(s-s') k(s-s')
  \label{eqn:energygaussnoise}
\end{align}
Defining $\Sigma^{2}_{p}(t)=\langle \langle
\hat{p}(t)\rangle^2\rangle_\xi - \langle \langle
\hat{p}(t)\rangle\rangle_\xi^2$ and similarly for the position
operator we find
\begin{align}
\Sigma^{2}_{p}(t) =  \Sigma^{2}_{x}(t) =  \epsilon(t).
\label{eq:variances_x_and_p}
\end{align}
We see that the variances of position and momentum with respect to
noise are controlled by the function $\epsilon(t)$ which is the energy
added to the system after stochastic forcing is turned on.  In
Section~\ref{sec:pdfs} we generalize these results and obtain all
moments of the noise-averaged position, momentum, and energy. The
complete distribution for position and momentum is Gaussian and
determined by its mean and variance. On the other hand, the
distribution for energy is exponential and thus characterized by its
mean and initial value. The noise-averaged energy of the oscillator
$\epsilon(t)$ appears frequently in our statistical analysis. 

If we consider Gaussian time-correlations, $\epsilon(t)$ has a closed
form in terms of error functions. However, to see its qualitative
properties, consider its derivatives. For the case of a Gaussian noise
(Eq.~\eqref{eq:6}),
\begin{align}
  \label{eq:15}
\frac{d \epsilon(t)}{d t } \xrightarrow{t\rightarrow\infty} \Lambda
\mathrm{e}^{-\omega^2\tau^2/2}.
\end{align}
 This means
that at long times the behavior is linear with slope $\Lambda
e^{-\omega^2\tau^2 /2}$.  The slope is exponentially small in $\tau$
with scale given by $1/\omega$. Thus, memory in the bath exponentially
suppresses the rate of energy transfer from bath to system at long
times.  From the second derivative
\begin{align}
  \label{eq:16}
 \frac{d^2 \epsilon(t)}{d t^2} =
\frac{2\Lambda}{\tau\sqrt{2\pi}} e^{-t^2/2\tau^2}\cos\omega t ,
\end{align}
we see that there are inflection points equally spaced in time which
means that at short times there are oscillations with fixed frequency
$\omega$ and their initial amplitude is of the order of $\Lambda$. It
also shows that the amplitude of such oscillations decay as time
increases with time scale $\tau$.  The
longer the memory of the noise the longer the oscillations are
prolonged. The short time oscillations and long time linear growth are
shown in Fig.~\ref{fig:2}. This behavior is generic to any noise
correlation function that decays fast enough. To understand this,
after a change of variables Eq.~\eqref{eqn:energygaussnoise} becomes
\begin{align}
  \label{eq:13}
  \epsilon(t) = t \int_{-\mathrlap{t}}^{\mathrlap{t}} du \, k(u) \cos\omega u
  - \int_{-\mathrlap{t}}^{\mathrlap{t}} du \, |u| k(u) \cos \omega u.
\end{align}
At long times $\epsilon(t)$ is linear and the first term in
Eq.~\eqref{eq:13} gives the slope of $\epsilon(t)$ as $t\rightarrow
\infty$ \cite{Maniscalco2004}. 

With these basic quantities defined and calculated, we can now find
the full quantum and statistical dynamics of the system characterized
by the density matrix and probability distribution functions.

\subsection{The noise-averaged density matrix}
\label{sec:noise-aver-dens}

The density matrix captures both the quantum and statistical nature of
a system, and in order to calculate it, we employ functional integral
and Lie algebraic methods illustrated in this section. 

The evolution operator for a single harmonic oscillator obeys the
equation $i \partial_t \hat U = \hat H (t) \hat U(t)$ with $ \hat U(0)
= \hat 1$, and is given by \cite{Galitski2011}
\begin{align}
  \hat U(t) = \mathrm{e}^{-i\omega t( \hat n + 1/2)} 
  \mathrm{e}^{-i(\Phi_1(t) \hat x + \Phi_2(t) \hat p)} \mathrm{e}^{i\gamma(t)}.
  \label{eq:9add}
\end{align}
where $\Phi_i(t) = \int_{0}^{t} ds \, \xi_j(s) R_{ji}(s)$. We define the 
noise-averaged density matrix by
\begin{align}
  \hat\rho(t) = \langle \hat U(t)\hat
  \rho(0) \hat U^{\dagger}(t)\rangle_\xi \equiv \mathrm{e}^{\mathcal{L}(t)}\hat\rho(0),
  \label{eq:liouvillian}
\end{align}
where $\hat \rho(0)$ is the initial density matrix. It is convenient
to express the evolution of the density matrix via a quantum
`Liouvillian' operator $\mathcal L(t)$.  Using $\mathrm{e}^{\hat A}
\hat B \mathrm{e}^{-\hat A} = \mathrm{e}^{\ad_{\hat A}}\hat B$ where
$\ad_{\hat A} \hat B =[\hat A,\hat B]$ is the (linear) adjoint
operator, we obtain
\begin{align}
  \label{eq:7}
  \mathrm{e}^{\mathcal L(t)} = \mathrm{e}^{-i\omega t \ad_{\hat n}}
  \braket{\mathrm{e}^{-i(\Phi_1(t) \adx + \Phi_2(t) \adp)}}_\xi.
\end{align}
Note that $[\adx,\adp]=0$ and $[\adx,1]=0=[\adp,1]$ allow us to treat
$\adx$ and $\adp$ as $c$-numbers when integrating over
$\xi_{1,2}$. Suppressing normalization, indices, and integration for
clarity, we obtain
\begin{align}
  \hat{\rho}(t) & = \mathrm{e}^{-i\omega t \ad_{\hat n}} \inbk{ \int
    \calD^2 \xi \; \mathrm{e}^{-i \int \bm \xi^T R\adX}
    \mathrm{e}^{-\frac12 \iint \bm \xi^T
      K^{-1}\bm \xi} }\hat \rho(0) \nonumber \\
  & = \mathrm{e}^{-i\omega t \ad_{\hat n}} \mathrm{e}^{-\frac12 \adX^T
    \inbk{\iint R^T K R} \adX }\hat \rho(0),\label{eq:71add}
\end{align}
where $\adX = (\adx,\adp)^T$.  In Eq.~\eqref{eq:71add}, we note that
the set of operators $\{ \frac12\adn, \adx^2+\adp^2, \adx^2-\adp^2,
2\adx\adp\}$ surprisingly form a Lie algebra (see
Appendix~\ref{sec:noise-algebra}).  This can be used to derive a full
equation of motion for the density matrix.  

Considering our particular form of noise, explicit calculation gives
\begin{align}
  \adX^T\inbk{\iint R^T K R }\adX& = \epsilon(t) (\adx^2 + \adp^2),
\end{align}
where $\epsilon(t)$ is given by Eq.~\eqref{eqn:energygaussnoise}.

To make further progress we need some facts about operators that act in this
Hilbert space. We know that any operator 
$\hat {\mathcal O}$ can be expanded (Appendix~\ref{sec:noise-algebra}) as
\begin{align}
  \label{eq:5}
  \hat{\mathcal O} = \int \frac{dy \, dq}{2\pi} \tr[\hat{ \mathcal
    O} \mathrm{e}^{iy\hat p - i q \hat x}] \mathrm{e}^{iq\hat x - i y \hat p},
\end{align}
and the operators $\mathrm{e}^{iq\hat x - i y \hat p}$ are
eigenoperators of the operators $\adx$ and $\adp$:
\begin{align}
  \adx \mathrm{e}^{iq\hat x - i y \hat p} & = y \, \mathrm{e}^{iq\hat x - i y \hat p},\label{eq:26}  \\
  \adp \mathrm{e}^{iq\hat x - i y \hat p} & = q \, \mathrm{e}^{iq\hat x - i y \hat p}. \label{eq:28}
\end{align}
Further, we can calculate the matrix element (Appendix~\ref{sec:eigenop_numberbasis})
\begin{align}
  \braket{n| \mathrm{e}^{iy \hat p - i q \hat x} | m} & =\sqrt{\frac{n!}{m!}}
  (z^*)^{m-n} L_n^{(m-n)}(|z|^2) \mathrm{e}^{-|z|^2/2}, \label{eq:79}
\end{align}
where $z \equiv (y+iq)/\sqrt{2}$ and $L_n^{(m)}$ is an associated
Laguerre polynomial. Also, $\adx^2+\adp^2$
commutes with $\adn$ (Appendix~\ref{sec:noise-algebra}).
The density matrix can be expanded as $\hat \rho(0) = \sum_{mn} \rho_{mn}
\ket m \bra n$ where $\rho_{mn}=\bra m \hat{\rho}(0) \ket n$ and therefore 
we only need to calculate the evolution
of the basis elements $\ket m \bra n$.  

Combining the above facts, we obtain from Eq.~\eqref{eq:5} that
\begin{multline}
 \mathrm{e}^{\mathcal L(t)} \ket m\bra n = \mathrm{e}^{-i\omega t(m-n)} \int \frac{d y \, d q}{2\pi}
  \braket{n | \mathrm{e}^{i y \hat p - i q \hat x} | m} \\ \times \mathrm{e}^{i q \hat x - i y \hat
  p} \mathrm{e}^{-\frac12 \epsilon(t)( q^2 + y^2) }.
\end{multline}
To evaluate this, we calculate the matrix element $\braket{k
  |\{\mathrm{e}^{\mathcal L(t)} \ket m\bra n \}| l }$; using
Eq.~\eqref{eq:79} and shifting to polar coordinates $z = \sqrt x
\mathrm{e}^{i\theta}$ such that $d^2 z = \frac12 dx \, d\theta$ we obtain
\newcommand{\melm}{ \braket{k| \{\mathrm{e}^{\mathcal L(t)} \ket m\bra n \}| l }}
\begin{multline}
 \melm   = \sqrt{\frac{n! l!}{m! k!}}
  \mathrm{e}^{-i\omega t(m-n)} \\ \times \int_{\mathrlap{0}}^{\mathrlap{\infty}} d x
  \int_{\mathrlap{0}}^{2\pi} \frac{d
    \theta }{2 \pi} x^{(k-l+m-n)/2} \mathrm{e}^{i\theta(k-l-m+n)}
  \\ \times L_n^{(m-n)}(x) L_{l}^{(k-l)}(x) \mathrm{e}^{-(1+\epsilon(t))x},\label{eq:10}
\end{multline}
for which we can integrate $\theta$ to obtain
\begin{multline}
  \melm= \sqrt{\frac{n! l!}{m!k!}}\delta_{k-l,m-n} \mathrm{e}^{-i\omega t(m-n)}
  \\ \times \int_{\mathrlap{0}}^{\mathrlap{\infty}} d x \, x^{m-n} L_n^{(m-n)}(x)
  L_{l}^{(m-n)}(x) \mathrm{e}^{-(1+\epsilon(t))x} .\label{eq:82}
\end{multline}

On the other hand, using the identity
\begin{align}
  \frac{(-x)^m}{m!}L_n^{(m-n)}(x) = \frac{(-x)^n}{n!} L_m^{(n-m)}(x),
\end{align}
we can rewrite Eq.~\eqref{eq:10} as
\begin{multline}
\melm  = \sqrt{\frac{m! k!}{n! l!}}\delta_{k-l,m-n} \mathrm{e}^{-i\omega t(m-n)}
  \\ \times \int_{\mathrlap{0}}^{\mathrlap{\infty}} d x \, x^{n-m} L_m^{(n-m)}(x)
  L_{k}^{(n-m)}(x) \mathrm{e}^{-(1+\epsilon(t))x}.\label{eq:83}
\end{multline}
The right hand side (RHS) of Eq.~(\ref{eq:83}) is the same expression
as the RHS of Eq.~(\ref{eq:82}) with $n \leftrightarrow m$ and $k
\leftrightarrow l$ (except for the multiplicative $\mathrm{e}^{-i\omega
  t(m-n)}$ term). Thus, we can use Eq.~\eqref{eq:82} and assume $m\geq
n$ without loss of generality. At the end of our calculation, we simply
switch indices to obtain $m \leq n$.

A change of variables $y = (1 + \epsilon(t)) x$ in Eq.~\eqref{eq:82}
yields
\begin{multline}
  \melm =\sqrt{\frac{n! l!}{k!  m!}}\frac{\delta_{k-l,m-n}\mathrm{e}^{-i\omega
      t(m-n)}}{(1+\epsilon(t))^{m-n+1}} \\
 \times  \int_{\mathrlap{0}}^{\mathrlap{\infty}} d y\, y^{m-n}
  L_n^{(m-n)}\inp{\tfrac y{1+\epsilon(t)}} L_{l}^{(m-n)}\inp{\tfrac
    y{1+\epsilon(t)}}\mathrm{e}^{-y}.\label{eq:84}
\end{multline}
Together with the property of Laguerre polynomials
\begin{align}
  L_n^{(m-n)}\inp{\tfrac y{1+\epsilon(t)}} = \sum_{i=0}^n
  \frac{\epsilon(t)^{n-i}}{(1+\epsilon(t))^n} \binom{m}{n-i}
  L_i^{(m-n)}(y),\label{eq:81}
\end{align}
we obtain 
\begin{multline}
  \int_{\mathrlap{0}}^{\mathrlap{\infty}} d y\, y^{m-n} \nonumber
   L_n^{(m-n)}\inp{\tfrac y{1+\epsilon(t)}} L_{l}^{(m-n)}\inp{\tfrac
    y{1+\epsilon(t)}}\mathrm{e}^{-y} \\
  =\frac{\epsilon(t)^{n+l}}{(1+\epsilon(t))^{n+l}}
  \binom{m}{n}\binom{l+m-n}{l}(m-n)! \\ \times
  \tensor[_2]{F}{_1}[-l,-n;1+m-n;\epsilon(t)^{-2}],
\end{multline}
where $\tensor[_2]{F}{_1}$ is the hypergeometric function. Thus, we
can return to Eq.~(\ref{eq:84}) to obtain $\melm$ and hence
$\mathrm{e}^{\mathcal L(t)}\ket m \bra n$ expanded in the number
basis. Then, given an arbitrary initial density matrix
\begin{align}
\hat \rho(0) = \sum_{mn} \rho_{mn} \ket m \bra n, 
\end{align}
we have the time-evolved, noise-averaged density matrix

\begin{widetext}
\begin{multline}
  \hat \rho(t) = \sum_{\mathclap{n,m=0}}^\infty \rho_{m+n,n}
  \mathrm{e}^{-i\omega tm} \sum_{l = 0}^\infty
  \sqrt{\frac{(m+n)!(l+m)!}{n!l!}}  \frac{\epsilon(t)^{n+l}
    \tensor[_2]{F}{_1}[-l,-n;1+m;\epsilon( t)^{-2}]}{m! (1 +
    \epsilon(t))^{m+n+l+1}} \ket{l+m}\bra{l} \\ +
  \sum_{\mathclap{\substack{ n=1\\m=0}}}^\infty \rho_{m,n+m}
  \mathrm{e}^{i\omega tn} \sum_{l = 0}^\infty
  \sqrt{\frac{(n+m)!(l+n)!}{m!l!}}  \frac{\epsilon(t)^{m+l}
    \tensor[_2]{F}{_1}[-l,-m;1+n;\epsilon( t)^{-2}]}{n!(1 +
    \epsilon(t))^{n+m+l+1}} \ket{l}\bra{l+n}.
    \label{eq:densitymatrix}
\end{multline}
\end{widetext}

\begin{figure}
  \centering
  \includegraphics[width=0.48\textwidth]{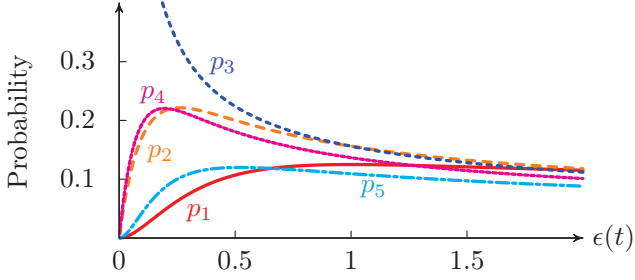}
  \caption{(Color online) The above is the [classical] probability
    $p_l$ of the oscillator being in state $\ket l$ if its initial
    state was $\ket 3$ vs.\ the oscillator energy $\epsilon(t)$
    [Eq.~\eqref{eq:example}].}
  \label{fig:occupation}
\end{figure}

Note that Eq.~(\ref{eq:densitymatrix}) only depends on time through
the energy added by noise $\epsilon(t)$ and a phase factor.  To get a
feeling for what Eq.~(\ref{eq:densitymatrix}) means consider an
oscillator in the initial state $\ket 3$, so $\hat\rho(0)=\ket3\bra3$
and 
\begin{align}
\hat \rho(t) &= \sum_l p_l[\epsilon(t)] \ket l \bra l 
\label{eq:example}
\end{align}
with $p_l[\epsilon(t)]$ being the probability of being in state $\ket
l$ at time $t$,
\begin{align}
  p_l[x]&= \tfrac{x^{l-3}[6x^6+18 l x^4 + 9 l (l-1)x^2 + l(l-1)(l-2)]
  }{6(1+x)^{l+4}}.
\label{eq:example2}
\end{align}
The results are plotted in Fig.~\ref{fig:occupation}. Since the
horizontal axis is $\epsilon(t)$, the non-Markovian oscillations at
short times can cause oscillations in the evolution of
$p_l[\epsilon(t)]$. This will be important when we consider
entanglement between two oscillators in
Section~\ref{sec:entanglement}.

The noise-average of any observable can be obtained by $\langle
\langle \hat X \rangle \rangle_\xi= \textrm{tr} \{ \hat \rho(t) \hat X
\}$ with $\hat \rho(t)$ given by Eq.~(\ref{eq:densitymatrix}). The
analytical computation of this quantity is one of our main results.
An alternative description which makes explicit the statistics of a
particular observable is via its probability distribution function
which we calculate in the next section.

\subsection{Probability distribution functions}
\label{sec:pdfs}

The random drive acting on the oscillator introduces uncertainty in
the quantum mechanical observables in addition to the quantum
mechanical spread of the system observables.  The effect of the random
drive on an observable can be completely characterized by its
probability distribution function (PDF).  For an
operator $\hat A$ with quantum mechanical average $\langle \hat
A\rangle$ its PDF, is
\begin{align}
P_{\hat A}[A;t] = \langle \delta[A-\braket{\hat A(t)}]\rangle_\xi.
\end{align} 
With this definition, we \emph{analytically} compute the PDF of
position, momentum, and energy with Gaussian noise.

\subsubsection{Position and momentum probability distribution functions}
\label{sec:pdfs_mom_pos}
\begin{figure}
  \includegraphics[width=0.48\textwidth]{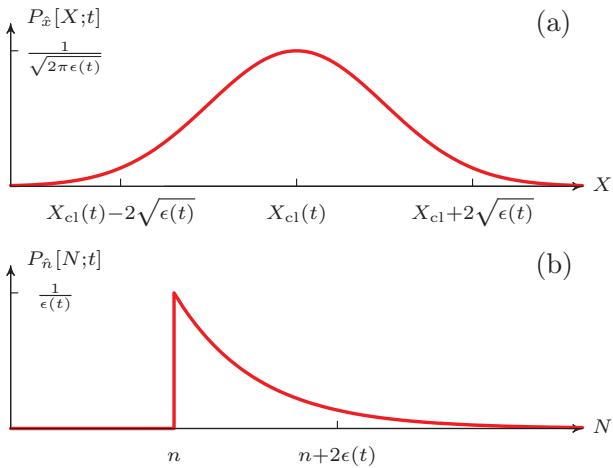} 
  \caption{(a) The probability distribution function for
    the position operator [Eq.~\eqref{eq:11}] is shown. The probability
    distribution function for momentum is the same except with a
    different center [$P_{\mathrm{cl}}(t)$ instead of
    $X_{\mathrm{cl}}(t)$]. (b) The probability distribution
    function of the energy or, equivalently, number operator
    [Eq.~\eqref{eq:pdfenergy}] is shown.}
  \label{fig:pdfs}
\end{figure}

The quantum mechanical average of position or momentum has the form 
\begin{align}
\braket{\hat A(t)} = X(t) + \int_{\mathrlap{0}}^{\mathrlap{t}}  d t'\, g_i(t,t') \xi_i(t').
\label{eq:15add}
\end{align}
For the moment we leave unspecified the functions $X(t)$ and
$g_i(t,t')$. The first term is the quantum mechanical average of the
operator in the absence of stochastic drive, e.g., from the first term
in Eq.~\eqref{eq:classical_eom} while the second term gives the
contribution due to the random drive.  This approach can be
applied to any operator that conforms to this form.  The Dirac delta
can be represented as an integral $\delta(x) = \int \mathrm{e}^{iu x}(
d u /2\pi)$ to obtain
\begin{multline}
  P_{\hat A}[A;t] = \int \frac{ d u }{2\pi} \int \calD^2
  \xi(t)\,  \mathrm{e}^{-\frac12\iint \bm \xi^T
    K^{-1}\bm \xi -  i u \int
    \mathbf{g}^T \bm \xi } \\ \times \mathrm{e}^{ i u (A- X(t))},
\end{multline}
where $\mathbf{g} = (g_1, g_2)^T$ and $\bm \xi = (\xi_1,\xi_2)^T$.
The above is a quadratic path integral and can thus be solved exactly
by standard techniques.  Throughout, we use the fact that
$K_{ij}(t,t') = K_{ji}(t',t)$. Explicit calculation gives
\begin{align}
  P_{\hat A}[A;t] &= \int\frac{ d u}{2\pi} \mathrm{e}^{-\frac{u^2} 2 \iint
    \mathbf{g}^T K \mathbf{ g} + i u
    (A-X(t))}\nonumber \\ & = \frac{1}{\sqrt{2\pi} \Sigma_A(t)}\exp\inbr{-\frac{
      [A-X(t)]^2}{2 \Sigma_A^2(t)} }\label{eq:27add},
\end{align}
where the variance of the PDF is
\begin{align}
  \Sigma_A^2(t) = \int_{\mathrlap{0}}^{\mathrlap{t}}  d t_1 \int_{\mathrlap{0}}^{\mathrlap{t}}  d t_2 \, g_i(t,t_1)
  K_{ij}(t_1,t_2) g_j(t,t_2).\label{eq:14add}
\end{align}
The form of $g_i$ depends on $\hat A$ (Eq.~(\ref{eq:15add})); for the
position operator $g_1(t,s)= - \sin\omega(t-s)$, $g_2(t,s)=
\cos\omega(t-s)$ and for the momentum operator $g_1(t,s)= -
\cos\omega(t-s)$, $g_2(t,s)= -\sin\omega(t-s)$ [these are taken directly
from the $R(t-s)\bm \xi$ term in Eq.~\eqref{eq:classical_eom}]. From
Eq.~\eqref{eq:27add} and given that the oscillator is in an initial
coherent state $\ket{z_0} $ where $z_0 =( x_0 + i p_0 )/\sqrt{2}$, we
obtain
\begin{align} 
   P_{\hat x}[X;t] & = \frac{1}{\sqrt{2\pi} \Sigma_x(t)}
   \exp\left\{-\frac{[X - X_{\mathrm{cl}}(t)]^2}{2 \Sigma_x(t)^2}\right\}, 
   \label{eq:11}\\
   P_{\hat p}[P;t] & = \frac{1}{\sqrt{2\pi}\Sigma_p(t) }
   \exp\left\{-\frac{[P - P_{\mathrm{cl}}(t)]^2}{2 \Sigma_p(t)^2}\right\},
   \label{eq:12}
\end{align} 
where the variances are the same as in Eq.~\eqref{eq:variances_x_and_p}, i.e.,
\begin{align}
  \Sigma^2_{x,p}(t) = \epsilon(t)
\label{eqn:sigma_pdf_new}
\end{align}
where $X,P$ are random variables of position and momentum. We see that
they are normally distributed about the solutions of the classical
equations of motion; $X_{\mathrm{cl}}(t) = x_0 \cos\omega t + p_0
\sin\omega t$ and $P_{\mathrm{cl}}(t) = - x_0 \sin\omega t + p_0
\cos\omega t$ (see Fig.~\ref{fig:pdfs}(a)).
Hence we see that $\langle\langle \hat x(t)\rangle\rangle_\xi$ and
$\langle\langle \hat p(t)\rangle\rangle_\xi$ will satisfy the standard
classical equations of motion for the harmonic oscillator.

It is important to note that memory effects are all included
analytically in Eq.~\eqref{eqn:sigma_pdf_new}. The spread in the
uncertainty is in fact in one-to-one correspondence with the behavior
of the energy of the oscillator.  This means that for white noise
($k(t) = \Lambda \delta(t)$) there is a Brownian (in time) increase in
the variance of the PDF of position and momentum, $\Sigma_{x,p}(t)
=\sqrt{\Lambda t}$, which is expected for a Markovian-type of
noise. If the memory of the noise is nonzero, the behavior is
non-monotonic and the uncertainty in the position and momentum can
decrease at times making the system more deterministic than
random. This is counter to what one might naively expect from a noise
source and shows the importance of memory. In the extreme
case of non-decaying noise correlations ($k(t) = \Lambda$), we have
$\Sigma^{2}_{x,p}(t) = 2 \Lambda (1-\cos\omega t)/\omega^2$ which
means that the PDFs $P_{\hat{x}}[X,2\pi k/\omega]$ and
$P_{\hat{p}}[P,2\pi k/\omega]$ (with $k$ an integer) collapse into
delta functions.  At these discrete times, the expectation value of
these observables will yield the classical value of position and
momentum and purely quantum mechanical behavior is restored. Thus,
even for finite, but long time-correlations, the position of $x$ can
stay localized for quite a long time (as seen by the exponential
suppression $e^{-\omega^2\tau^2/2}$ of the growth of the variance).
Intuitively, the system remembers its initial pure state and tries to
restore it.  When the memory is finite, this restoration is not
complete but still can give non-monotonic behavior.  Nonetheless, at
large times we recover Brownian-type behavior (see Fig.~\ref{fig:2}).

\subsubsection{Energy probability distribution function}
\label{sec:pdfs_energy}
Formally consider the quantity
\begin{align}
  \braket{\hat B(t)} = X(t) + \frac12\int_{\mathrlap{0}}^{\mathrlap{t}}  d t_1\int_{\mathrlap{0}}^{\mathrlap{t}}  d t_2 \,  \xi_i(t_1) F_{ij}(t_1,t_2) \xi_j(t_2),
\end{align}
where the function $X(t)$ and matrix $F(t)$ are unspecified for the moment. 
This is the case for the quantum expectation value of the energy. 
In this case explicit calculation gives,
\begin{multline}
  P_{\hat B}[B; t]  = \int \frac{ d u}{2\pi}\frac1{\sqrt{\det{K}}}
  \int \calD^2 \xi(t)\, \\ \times \mathrm{e}^{-\frac12 \iint \bm
    \xi^T\inbk{K^{-1}+  i u
      F} \bm \xi +  i u (B-X(t))} 
\end{multline}
Evaluating this quadratic path integral yields
\begin{align}
P_{\hat B}[B;t]=\int \frac{ d u}{2\pi}
  \frac{\mathrm{e}^{ i u (B-X(t))}}{\sqrt{\det[1 + i u K F]}}.\label{eq:22add}
\end{align}
The determinants can be viewed in the following way: To find ``$\det
D(t_1,t_2)$'' take the function $D(t_1,t_2)$ and time slice it $N-1$
times from $0$ to $t$, so one has an $N\times N$ matrix. Find the
determinant of this matrix, then let $N\rightarrow \infty$. The
quantities $1+iuKF$
and $K$ are also $2\times 2$ matrices, and in that case one just takes
the determinant of the matrix created by the direct product of those
two spaces ($2N \times 2N$ matrices).  Again, we use the fact that
$K_{ij}(t_1,t_2) = K_{ji}(t_2,t_1)$ and further, we assume
$F_{ij}(t_1,t_2) = F_{ji}(t_2,t_1)$.

For the specific case of the energy PDF we compute this determinant 
using methods developed in Section~\ref{sec:noise-aver-dens}. Assuming 
the system is initially in a number state $\ket n$, the average occupation 
number in the presence of the external drive is
\begin{align}
  \label{eq:11add}
  \braket{\hat n(t)} = n + \frac12\int_{\mathrlap{0}}^{\mathrlap{t}} d t_1 \int_{\mathrlap{0}}^{\mathrlap{t}} d t_2\, \xi_i(t_1)
  R_{ij}(t_1-t_2) \xi_j(t_2),
\end{align}
and using Eq.~\eqref{eq:22add} 
\begin{align}
  \label{eq:12add}
 P_{\hat n}[N;t] =\int \frac{ d u}{2\pi}\sqrt{\frac{1}{\det[1 + 
      i u K R]}} \, \mathrm{e}^{ i u (N-n)}.
\end{align}
By going a step back, the determinant can be written as
\begin{align}
  \label{eq:16add}
\sqrt{\frac{1}{\det[1 +  i u K R]}} =   \braket{\mathrm{e}^{-\frac12 i u \xi^T R \xi}}_{\xi} .
\end{align}
Now, let us consider the following quantity
\begin{multline}
  \label{eq:17add}
\braket{ \braket{0|\{ {\mathrm{e}^{-iz(\Phi_1(t) \adx + \Phi_2(t) \adp) }}
    \ket 0 \bra 0\}|0}}_{\xi} \\=
 \braket{0|\{ \braket{\mathrm{e}^{-iz(\Phi_1(t) \adx + \Phi_2(t) \adp) }}_{\xi} \ket 0 \bra 0\}|0},
\end{multline}
The left hand side of of Eq.~\eqref{eq:17add} can be found by standard
techniques
\begin{multline}
  \label{eq:19add}
\braket{ \braket{0|\{ {\mathrm{e}^{-iz(\Phi_1(t) \adx + \Phi_2(t) \adp) }}
    \ket 0 \bra 0\}|0}}_{\xi} \\=  \braket{\mathrm{e}^{-\frac12 z^2 \xi^T R \xi}}_{\xi}.
\end{multline}
However, the RHS of Eq.~\eqref{eq:17add} can be calculated just as in
Section~\ref{sec:noise-aver-dens} (see Eq.~\eqref{eq:7}) with
$\Phi_i(t)\rightarrow z \Phi_i(t)$ or equivalently
$\epsilon(t)\rightarrow z^2 \epsilon(t)$.  The RHS of
Eq.~\eqref{eq:17add} is then the same as
letting $\hat \rho_z(0) = \ket 0 \bra 0$ and evaluating
$\braket{0|\hat \rho_z(t)|0}$ with the suggested substitutions (the
subscript $z$ represents this substitution). Using
Eq.~\eqref{eq:densitymatrix}, we have
\begin{align}
  \label{eq:9}
  \hat \rho_z(t) = \sum_{l=0}^\infty \frac{z^{2l} \epsilon(t)^l}{[1+
    z^2 \epsilon(t)]^{l+1}} \ket l \bra l.
\end{align}
Reading off the $\ket 0 \bra
0$ component in Eq.~\eqref{eq:9}, we obtain
\begin{align}
  \label{eq:18add}
  \braket{0|\{ \braket{\mathrm{e}^{-iz(\Phi_1(t) \adx + \Phi_2(t) \adp) }}_{\xi} \ket 0 \bra
    0\}|0} =\frac1{1+z^2\epsilon(t)}.
\end{align}
Letting $z^2 =  i u$, we get the identity
\begin{align}
  \label{eq:20add}
   \sqrt{\frac{1}{\det[1 +  i u K R]}} =\frac1{1+ i u \epsilon(t)}.
\end{align}
The PDF for the number operator is then 
\begin{align}
  \label{eq:21add}
 P_{\hat n}[N;t] = \int \frac{ d u}{2\pi}\frac{\mathrm{e}^{ i u (N-n)}}{1+ i
    u \epsilon(t)}.
\end{align}
Since $\epsilon(t)>0$, this quantity is non-zero if $N>n$, 
and is calculated with contour integration. We get a quantity
that is only implicitly dependent on time through $\epsilon(t)$,
\begin{align} 
 P_{\hat{n}}[N;t] = \frac{ \Theta(N- n )}{\epsilon(t)} \mathrm{e}^{-(N-n)/\epsilon(t)},
  \label{eq:pdfenergy}
\end{align}
where $N$ is the continuous number random variable and 
$\Theta$ is the step function; see Fig.~\ref{fig:pdfs}(b). Eq.~\eqref{eq:pdfenergy} implies
that the energy will never statistically fluctuate lower than the
initial value. The exponential PDF has mean $n + \epsilon(t)$ and
variance $\epsilon(t)^2$ with the memory of the noise entering only 
via $\epsilon(t)$; see Fig.~\ref{fig:2}. The non-Markovian
effects will cause the PDF to narrow as well, and in the limit of
infinite noise correlation-time it will periodically return to
$\delta(N-n)$ just as in the case for position and momentum.

\section{Entanglement dynamics}
\label{sec:entanglement}

Having computed the density matrix for a single oscillator in the
presence of a random non-Markovian drive, we are in position to
study entanglement dynamics in an exact manner. We extend the solution
to two independent oscillators (each with its own independent,
stochastic, non-Markovian drive) initially in an entangled state. The
goal is to characterize how the entanglement evolves in time and in
particular the effects of the memory of the noise on the entanglement
dynamics. Extending the notation of Section~\ref{eq:singleho}, the
Hamiltonian for the driven oscillators is
\begin{multline}
  \hat H = \omega(a^\dagger a + \tfrac12) + \tfrac1{\sqrt2}[\xi (t) a^\dagger + \mathrm{h.c.}] \\
  +\omega(b^\dagger b + \tfrac12) + \tfrac1{\sqrt2}[\eta (t) b^\dagger
  + \mathrm{h.c.}].
\label{eq:hamiltonian_2osc}
\end{multline}
where $\xi,\eta$ are the stochastic fields (both have the same
statistics but are independent of one another). Using
Eq.~\eqref{eq:liouvillian}, the evolution of the \emph{two-oscillator}
density matrix $\hat{\varrho}(t)$ is given by
\begin{align}
\hat \varrho(t) = \mathrm{e}^{\mathcal{L}_1(t)}\otimes \mathrm{e}^{\mathcal{L}_2(t)} \hat{\varrho}(0) ,
\end{align}
with the initial density matrix corresponding to a maximally entangled
state in the two lowest levels of the oscillators
\begin{align}
 \hat{\varrho}(0)= \frac{1}{2}(\ket{01} + \ket{10}) \otimes (\bra{01} + \bra{10}),
  \label{eq:4} 
\end{align}
where $\ket{nm}$ represents the first oscillator in state $\ket{n}$
and the second in state $\ket{m}$. We can apply
Eq.~\eqref{eq:densitymatrix} to each of the states $\ket{0}\bra{0}$,
$\ket{0}\bra{1}$, $\ket{1}\bra{0}$, and $\ket{1}\bra{1}$ separately.
The density matrix can then be written as $\hat{\varrho}(t)=
\sum_{nm,n'm'} \braket{nm| \hat{\varrho}(t) |n'm'}
\ket{nm}\bra{n'm'}$. But we are only interested in how the qubit-like
entanglement in the subspace $\{ \ket{00}, \ket{01}, \ket{10},
\ket{11}\}$ evolves in time.  This defines a new $4\times 4$ density
matrix $\hat{\varrho}_2$ given by $ \Pi \hat{\varrho}(t) \Pi$ where $
\Pi =\sum_{n,m=0}^1 \ket{nm}\bra{nm}$ is the projection operator onto
the subspace. We normalize this expression by the trace of $\Pi
\hat{\varrho}(t) \Pi$ for convenience, but this does not affect our
conclusions. Explicit calculation gives
\begin{align}
   \hat{\varrho}_2 = 
  \begin{pmatrix} \frac{\epsilon(t)}{[1+ \epsilon(t)]^3}& 0 & 0 & 0 \\
    0 & \frac{\frac12 + \epsilon(t)^2}{[1+ \epsilon(t)]^4} &
    \frac{1/2}{[1+\epsilon(t)]^4} & 0 \\ 0 & \frac{1/2}{[1+
      \epsilon(t)]^4} & \frac{\frac12 + \epsilon(t)^2}{[1+\epsilon(t)]^4}
    & 0 \\ 0 & 0 & 0 & \frac{\epsilon(t)[1+\epsilon(t)^2]}{[1+
      \epsilon(t)]^5}
  \end{pmatrix}.
 \label{eq:8} 
\end{align}
Given this density matrix we compute the concurrence as given in
Eq.~\eqref{eq:concurrencedef}. The results are presented in
Fig.~\ref{fig:energy-concurrence-plot-final.eps} along side plots of
the energy of a \emph{single} oscillator. To see the connection
between concurrence and energy, it can be shown that the energy given
to a single oscillator by the stochastic field is again
Eq.~\eqref{eqn:energygaussnoise} (more precisely: the energy is the
average of the energies for $\ket 0$ and $\ket 1$ time evolved
separately). Since this $\hat\varrho_2$ only explicitly depends on the energy
$\epsilon(t)$, $\epsilon(t)$ effectively controls the entanglement.
In Fig.~\ref{fig:energy-concurrence-plot-final.eps} we show the
behavior of the energy and concurrence for different noise correlation
times.  We see that for white noise the energy increases linearly as a
function of time and the concurrence vanishes at a critical time
$\Lambda t_\mathrm{c} \approx 0.455$.  For noise with memory,
non-Markovian oscillations of the energy lead to sudden death
(rebirth) of the entanglement as the energy crosses above (below) a
specific initial-state dependent threshold $\epsilon_\mathrm{c}\approx
0.455$. Intuitively, the system `remembers' it quantum state,
particularly its entanglement. This rebirth phenomenon is absent in
baths with no memory
(Fig.~\ref{fig:energy-concurrence-plot-final.eps}(a,b)).

In terms of our initial density matrix, we may generate entanglement
between higher energy states. Letting $P = 1 - \Pi$ be the projection
onto the rest of the Hilbert space, then the density matrix can be
decomposed as $\hat{\varrho} = \Pi \hat{\varrho} \Pi + P \hat{\varrho}
\Pi + \Pi \hat{\varrho} P + P \hat{\varrho} P$, and only the first
term $ \Pi \hat{\varrho} \Pi$ is separable when $C(t)=0$ (precisely:
it can be written as the sum of density matrices of separable states)
while the higher energy states may still exhibit entanglement between
themselves and the lower energy states.  Intuitively, the higher
energy states act as a ``cavity'' to their respective ``qubit'' (as in
\cite{Yonac2006,*Yonac2007}), so one may expect entanglement is being
transferred back and forth between them (as the classical noise slowly
diminishes the overall entanglement).

\section{Summary}
\label{sec:conclusions}

\begin{table*}[]
 \renewcommand{\arraystretch}{2}
\renewcommand{\tabularxcolumn}[1]{>{\arraybackslash}m{#1}}
\newcolumntype{Y}{>{\raggedright}X}
\begin{tabularx}{\textwidth}{>{\raggedright}m{2cm}m{5pt}Y>{\raggedright}m{2.5cm}lm{1.5cm}}
  Quantity && Noise-averaged expression & Initial state && Reference \\
  \hline Energy added by noise && $\displaystyle \omega\epsilon(t) =
  \omega \int_{\mathrlap{0}}^{\mathrlap{t}}ds
  \int_{\mathrlap{0}}^{\mathrlap{t}} ds'\, \cos\omega(s-s') k(s-s')$ &
  Any && Eq.~\eqref{eqn:energygaussnoise}  \\
  Density matrix && $ \displaystyle \hat \rho(t) =
  \sum_{\mathclap{n,m,l}}\rho_{m+n,n} \mathrm{e}^{-i\omega tm}
  {\scriptstyle \sqrt{\frac{(m+n)!(l+m)!}{n!l!}}  \frac{\epsilon(t)^{n+l}
      \tensor[_{2}]{F}{_1}[-l,-n;1+m;\epsilon( t)^{-2}]}{m! (1 +
      \epsilon(t))^{m+n+l+1}} }\ket{l+m}\bra{l} $
  $\displaystyle\phantom{\hat\rho(t) = l} + \sum_{\mathclap{n>0,m,l}}
  \rho_{m,n+m} \mathrm{e}^{i\omega tn}
  {\scriptstyle\sqrt{\frac{(n+m)!(l+n)!}{m!l!}}  \frac{\epsilon(t)^{m+l}
      \tensor[_2]{F}{_1}[-l,-m;1+n;\epsilon( t)^{-2}]}{n!(1 +
      \epsilon(t))^{n+m+l+1}}} \ket{l}\bra{l+n}$ &
  $\displaystyle\sum_{nm}\rho_{mn}\ket n \bra m$ &&
  Eq.~\eqref{eq:densitymatrix} \\
  Position PDF && $\displaystyle P_{\hat x}[X;t] =
  \frac{1}{\sqrt{2\pi\epsilon(t)} } \exp\left\{-\frac{[X -
      X_{\mathrm{cl}}(t)]^2}{2 \epsilon(t)}\right\}$ & $e^{z_0
    a^\dagger - z_0^* a} \ket 0$
  && Eq.~\eqref{eq:11} \\
  Momentum PDF && $\displaystyle P_{\hat p}[P;t] =
  \frac{1}{\sqrt{2\pi\epsilon(t)}} \exp\left\{-\frac{[P -
      P_{\mathrm{cl}}(t)]^2}{2 \epsilon(t)}\right\}$ & $e^{z_0
    a^\dagger - z_0^* a} \ket 0$
  && Eq.~\eqref{eq:12} \\
  Energy PDF && $\displaystyle P_{\hat{n}}[N;t] = \frac{ \Theta(N- n
    )}{\epsilon(t)} \mathrm{e}^{-(N-n)/\epsilon(t)}$ & $\ket n$ &&
  Eq.~\eqref{eq:pdfenergy} \\ \\[-18pt]
  Two oscillator density matrix && $\hat{\varrho}(t) = \bordermatrix{
    &{\color{mygray} \bra{00}} & {\color{mygray} \bra{01}} &
    {\color{mygray} \bra{10}} & {\color{mygray} \bra{11}} & {\color{mygray} \cdots} \cr
   {\color{mygray} \ket{00}} & \frac{\epsilon(t)}{[1+
      \epsilon(t)]^3}& 0 & 0 & 0 & \cdots \cr 
{\color{mygray} \ket{01}}  & 0 & \frac{\frac12 +
      \epsilon(t)^2}{[1+ \epsilon(t)]^4} &
    \frac{1/2}{[1+\epsilon(t)]^4} & 0 & \cr
 {\color{mygray} \ket{10}} & 0 & \frac{1/2}{[1+
      \epsilon(t)]^4} & \frac{\frac12 +
      \epsilon(t)^2}{[1+\epsilon(t)]^4} & 0 \cr 
{\color{mygray} \ket{11}}  &0 & 0 & 0 &
    \frac{\epsilon(t)[1+\epsilon(t)^2]}{[1+ \epsilon(t)]^5}\cr {\color{mygray} \vdots}
    & \vdots & & & &\ddots } $ & $\tfrac1{\sqrt2}(\ket{01}+\ket{10})$
  && Eq.~\eqref{eq:8}
  \end{tabularx}

  \caption{Summary of results. The individual harmonic oscillator 
    Hamiltonian is $\hat H = \omega (a^\dagger a +1/2) + \xi_1(t) \hat x + \xi_2(t)
    \hat p$ where $\xi_{1,2}(t)$ are stochastic fields satisfying
    $\braket{\xi_i(t)}_\xi = 0$,
    $\braket{\xi_i(t)\xi_j(t')}_\xi = \delta_{ij}k(t-t')$, and Gaussian
    distributed. Probability distribution functions (PDFs) are defined
    by $P_{\hat A}[A;t] = \braket{\delta(A - \braket{\hat
        A})}_\xi$. The complex number $z_0$ defines both a coherent
    state and the point in
    phase space where the classical solutions to the harmonic
    oscillator, $X_{\mathrm{cl}}(t)$ and $P_{\mathrm{cl}}(t)$,
    begin. The elipses in the two-oscillator density matrix
    represent quantities not explicitly calculated in text.}
\label{table:summary}
\end{table*}

In summary, we developed Lie algebraic and functional methods to
\emph{analytically} study the statistics of a single oscillator in the
presence of stochastic drive with memory; see
Table~\ref{table:summary} for our analytical results.  We found
analytical expressions for the density matrix
(Eq.~\eqref{eq:densitymatrix}) and the probability distribution
functions of position, momentum, and energy (Eq.~\eqref{eq:11},
Eq.~\eqref{eq:12} and Eq.~\eqref{eq:pdfenergy} respectively). These
expressions fully capture the statistics of the observables and
explicitly show that the uncertainty can decrease at times in a
non-Markovian environment. In all of these expressions we saw that
memory effects are encoded in the noise-averaged energy. 

Calculating the noise-averaged energy, we found a non-monotonic
behavior for sufficiently long time-correlations in the bath. This
non-monotonic behavior controls many things throughout, including the
death and subsequent rebirth of entanglement for two uncoupled
oscillators considered in Section \ref{sec:entanglement} and the
variance in the position, momentum, and energy PDFs.  Diffusive
behavior is established at times much longer than $\tau$; in this
regime the energy (and variances of position and momentum) is linear
in time with a slope that decreases exponentially as $\tau$ increases,
$\mathrm{e}^{-\omega^2\tau^2/2}$; Fig.~\ref{fig:2}. The suppression of
the slope also implies that the position can remain localized to a
small region in real space when $\tau$ is large, as seen explicitly in the
position probability distribution function.

The position and momentum PDFs are normally distributed
about their classical trajectories in the absence of a drive
(Eq.~\eqref{eq:11} and Eq.~\eqref{eq:12}).  Interestingly, memory
effects enter \emph{only} through the energy added to the system,
Fig.~\ref{fig:2}. Thus, non-monotonic behavior of the energy implies
non-monotonic behavior of the variance in position and
momentum -- i.e., variance can decrease for times shorter than
$\tau$.
On the other hand, the PDF for energy is
exponential (Eq.~\eqref{eq:pdfenergy}) with mean 
$n+\epsilon(t)$, i.e., proportional to the energy of the oscillator
($n$ is the initial number state). We find, again, that the memory
effects enter only via the energy and hence similar oscillations of
the width of the energy PDF are predicted.

Using functional integral and Lie algebraic methods, we also found
an analytical expression for the density matrix
(Eq.~\eqref{eq:densitymatrix} and Table~\ref{table:summary}), and
interestingly, we again found that all memory effects enter only
through the energy $\epsilon(t)$. We used this expression to find the
concurrence in the two lowest lying states of two independent
oscillators. Just as the density matrix only depended on
$\epsilon(t)$, so too did the concurrence. Therefore, the non-monotonic
behavior in energy for correlated noise implies non-monotonic behavior
for concurrence.  This is the origin of the oscillations in the
concurrence seen in
Fig.~\ref{fig:energy-concurrence-plot-final.eps}. In particular, there
is a threshold of energy above which the oscillators disentangle
completely but below which they remain entangled. Hence the sudden
death and rebirth of entanglement are due to the energy of single
oscillators crossing this threshold back and forth
(Fig.~\ref{fig:energy-concurrence-plot-final.eps}). These oscillations
in turn are due to the effects of the memory in the noise. Physically,
the higher energy states in each oscillator act as the ``cavity'' to
their respective ``qubit'' (composed of the two lowest lying states),
potentially storing the entanglement as the classical noise slowly
kills off entanglement entirely.

Nano-mechanical oscillators could provide a possible experimental
realization of some of the effects studied in this work. While usually
interacting with an environment that is highly fluctuating can cause
the oscillators to behave classically, recent experiments have been
able to cool them to their ground state and excite either a
single quanta of energy or coherent state \cite{OConnell2010}. These
systems have applications ranging from fundamental research to mass
sensors, and understanding the effects of noise on the dynamics of
entanglement on such objects has potential technological applications
\cite{Jensen2008}.

To conclude, every quantity calculated shows that the system
``remembers'' its quantum state, and given a long bath memory, the
system can partially restore its quantum state for short intervals of
time -- even if that means restoring entanglement after its
destruction.



 \section*{Acknowledgments} 

 We thank Sankar Das Sarma, Lev S. Bishop, and Edwin Barnes for their
 comments. This research was supported by NSF-CAREER award (VG and JW)
 and JQI-PFC (BF and VG).

\appendix

\section{The noise algebra}
\label{sec:noise-algebra}

In Section~\ref{sec:noise-aver-dens} we found a set of operators which
we claim is a Lie algebra:
\begin{align*}
  \mathfrak K \equiv \operatorname{\mathrm{span}}\{ \tfrac12\adn\, , \, \adx^2 - \adp^2\, ,\, 2 \adx \adp\, ,\, \adx^2 + \adp^2\} .
\end{align*}
To see this, we need to consider commutators. We introduce the general
operator $\hat{\mathcal O}$ to act as an operator which these adjoint
operators act on. First, we establish that $[\adx, \adp] =0$ by the
Jacobi identity
\begin{align}
  [\adx, \adp]\hat{\mathcal O} & = [x,[p,\hat{\mathcal O}]] - [p, [x,
  \hat{\mathcal O}]] \\ & = [[x,p],\hat{\mathcal O}] = i [1, \hat{\mathcal O}] = 0.
\end{align}
This establishes that
\begin{align}
    [\adx^2+ \adp^2, 2\adx\adp] & = 0,\\
    [\adx^2+ \adp^2, \adx^2 - \adp^2] & = 0,\\
    [2\adx\adp, \adx^2 - \adp^2] & = 0.
\end{align}
The interesting pieces then come from the evaluation of
\begin{align*}
  [\tfrac12\adn, \adx^2]\hat{\mathcal O} & 
  = -i \adx\adp \hat{\mathcal O},\\
  [\tfrac12\adn,\adp^2] \hat{\mathcal O} & = i \adx\adp \hat{\mathcal O}, \\
  [\tfrac12\adn, \adx\adp ]\hat{ \mathcal O}
  & 
= i(\adx^2 - \adp^2) \hat{\mathcal O}.
\end{align*}
And we get the commutators
\begin{align}
  [\tfrac12\adn, \, \adx^2 + \adp^2] &= 0 \\
  [\tfrac12\adn, \, \adx^2 - \adp^2] &= -2 i \adx \adp \\
  [\tfrac12\adn, \, 2\adx\adp] & = i(\adx^2-\adp^2). 
\end{align}
Since aside from $\tfrac12\adn$, each operator commutes with each
other and can be simultaneously diagonalized. In fact, we can write
the complete set of eigenoperators that simultaneously diagonalize
$\adx^2+ \adp^2$, $\adx^2 - \adp^2$, and $2\adx\adp$ as $\mathrm{e}^{iq\hat
  x-i y \hat p}$.  This comes from 
\begin{align}
  \adx \mathrm{e}^{-iy\hat p} & = y \,\mathrm{e}^{-iy\hat p}, & \adp
  \mathrm{e}^{iq \hat x} = q \, \mathrm{e}^{i
    q \hat x}.\label{eq:69appendix}
\end{align}
The eigenvalues of our Lie algebraic generators are then
\begin{align}
  [\adx^2 + \adp^2]\mathrm{e}^{iq\hat x-i y \hat p} & = [y^2 + q^2]\mathrm{e}^{iq\hat
    x-i y\hat p}, \\
  [\adx^2 - \adp^2]\mathrm{e}^{iq\hat x-i y \hat p} & = [y^2 - q^2]\mathrm{e}^{iq\hat
    x-i y \hat p}, \\
  2\adx\adp \mathrm{e}^{iq\hat x-i y \hat p} & = 2 q y \, \mathrm{e}^{iq\hat x-i y
    \hat p}.
\end{align}

Furthermore, these eigenoperators are \emph{complete} (i.e.\ any operator
in this space is a linear combination of them). In order to see this,
we can take any operator and write it in terms of the already complete
basis $\ket x \bra {x'}$ as follows
\begin{align}
  \hat{ \mathcal O }& = \int d x\,  d x'\,  \mathcal O(x,x') \ket x \bra{x'}\label{eq:70appendix}
\end{align}
Now, we change variables from $(x, x')$ to $(x,y)$ where
$y = x - x'$ so that we can write $\ket x \bra{x'} = \ket x \bra{x-y}
= \ket x \bra x \mathrm{e}^{-i y \hat p}$. Then, focusing on $\ket
x \bra x$ and recall that $\braket{x|p} = \mathrm{e}^{ixp}$, we get
\begin{align}
  \ket x \bra x = \int\frac{ d p\,  d p'}{(2\pi)^2}
  \mathrm{e}^{-ix(p-p')}\ket p \bra{p'}.
\end{align}
We can let $p = p' + q$ to obtain
\begin{align}
  \ket x \bra x = \int\frac{ d q\,  d p'}{(2\pi)^2} \mathrm{e}^{-ix q}\mathrm{e}^{iq
    \hat x} \ket {p'} \bra{p'} = \int \frac{ d q}{2\pi} \mathrm{e}^{-ix q}\mathrm{e}^{iq \hat x}.
\end{align}
Plugging back into Eq.~(\ref{eq:70appendix}), we obtain
\begin{align}
 \hat{ \mathcal O }& = \int  \frac{d y d q}{2\pi}\,
  \inbk{\int d x \, \mathcal O(x,x-y) \mathrm{e}^{-ix q}}\mathrm{e}^{iq \hat x} \mathrm{e}^{-iy
    \hat p}.\label{eq:72appendix} 
\end{align}
Thus, any operator can be written as a linear combination of $\mathrm{e}^{iq
  \hat x} \mathrm{e}^{-iy \hat p}$ and hence also $\mathrm{e}^{iq
  \hat x-iy \hat p}$. Using the fact that $\mathcal
O(x,x') = \braket{x|\hat{ \mathcal O}| x'}$, we can evaluate
Eq.~(\ref{eq:72appendix}) one step further
\begin{align}
  \hat{ \mathcal O }& = \int   \frac{d y d q}{2\pi}\,
  \tr\inbk{\hat{\mathcal O} \mathrm{e}^{iy\hat p-iq\hat x}}\mathrm{e}^{iq \hat x-iy
    \hat p}.\label{eq:75appendix}
\end{align}

As a Lie algebra, using Baker-Campbell-Hausdorf relations, one can
obtain, from Eq.~\eqref{eq:71add}, an equation of motion for $\hat
\rho(t)$ of the form
\begin{multline}
  \label{eq:14}
  i \partial_t \hat \rho(t) = [\omega \adn + \alpha_1(t)\, 2\adx \adp
  \\+ \alpha_2(t) (\adx^2 - \adp^2) + \alpha_3(t) (\adx^2 + \adp^2)]
  \hat \rho(t).
\end{multline}

\section{Eigenoperators $\mathrm{e}^{i y \hat p - i q \hat x}$ in the number basis}
\label{sec:eigenop_numberbasis}

We compute the matrix element $\braket{ n |\mathrm{e}^{i y \hat p - i q
  \hat x}| m}$.  We can rewrite $\mathrm{e}^{i y \hat p - i q \hat
  x}= \mathrm{e}^{-z a^{\dagger}}\mathrm{e}^{z^{*}
  a}\mathrm{e}^{-|z|^2/2}$ where $z= (y-i q)/\sqrt{2}$ then explicit
calculation gives
\begin{multline}
\braket{ n |\mathrm{e}^{i y \hat p - i q \hat x}| m}\\ =
\frac{1}{\sqrt{n! m!}} \braket{ 0 |a^{n}\mathrm{e}^{-z a^{\dagger}}\mathrm{e}^{z^{*} a}(a^\dagger)^{m}| 0} \mathrm{e}^{-|z|^2/2}. \label{eq:1}
\end{multline}
Inserting $1= \mathrm{e}^{-z^{*} a} \mathrm{e}^{z^{*} a}$ a total of
$m$ times, we find $\mathrm{e}^{z^{*} a}(a^\dagger)^{m} = (a^{\dagger}
+ z^{*})^{m}\mathrm{e}^{z^{*} a}$ and a similar manipulation gives
$a^n \mathrm{e}^{-z a^{\dagger}} = \mathrm{e}^{-z a^{\dagger}}(a -
z)^{n}$.  Substituting back into Eq.~\eqref{eq:1}, we obtain
\begin{align}
  \braket{ n| \mathrm{e}^{i y \hat p - i q \hat x}|m} \hspace{-35pt} &
 \nonumber \\ & = \frac1{\sqrt{n!m!}}\braket{ 0|
    (a -z)^{n} (a^{\dagger} + z^{*})^m |0}
  \mathrm{e}^{-|z|^2/2} \nonumber \\
  & = \sum_{i=0}^{n}\sum_{j=0}^{m} \sqrt{\frac{n!m!}{i!j!}}\frac{
    \braket{i | j} (-z)^{n-i}(z^{*})^{m-j}}{(n-i)!
    (m-j)!}\mathrm{e}^{-|z|^2/2}  \nonumber\\
  & = \sqrt{\frac{n!}{ m!}} \sum_{i=0}^{{\mathrm{min}(n,m)}} \frac{ m! (-z)^{n-i}(z^{*})^{m-j}}{i! (n-i)! (m-j)!}\mathrm{e}^{-|z|^2/2}  \nonumber \\
  & = \sqrt{\frac{n!}{m!}} (z^{*})^{m-n} L_{n}^{(m-n)}(|z|^2)
  \mathrm{e}^{-|z|^2/2},
\end{align}
where $L_{n}^{(m)}$ are the generalized Laguerre polynomials.

\bibliography{ho}

\end{document}